\documentclass[english,aps,preprint]{revtex4}
\usepackage[T1]{fontenc}
\usepackage[latin9]{inputenc}
\usepackage{amsmath}
\usepackage{amssymb}
\usepackage{graphicx}
\usepackage{esint}

\makeatletter
\@ifundefined{textcolor}{}
{%
 \definecolor{BLACK}{gray}{0}
 \definecolor{WHITE}{gray}{1}
 \definecolor{RED}{rgb}{1,0,0}
 \definecolor{GREEN}{rgb}{0,1,0}
 \definecolor{BLUE}{rgb}{0,0,1}
 \definecolor{CYAN}{cmyk}{1,0,0,0}
 \definecolor{MAGENTA}{cmyk}{0,1,0,0}
 \definecolor{YELLOW}{cmyk}{0,0,1,0}
}

\usepackage{amsmath}
\usepackage{bbm}

\makeatother

\usepackage{babel}
\begin{document}

\preprint{BROWN HET-1632}

\title{\textmd{Loop effects and infrared divergences in slow-roll inflation}}

\author{\textup{Klaus Larjo}}

\email{klaus.larjo@gmail.com}

\affiliation{Department of Physics, Brown University, Providence, RI, 02912, USA}

\author{\textup{David A. Lowe}}

\email{lowe@brown.edu}

\affiliation{Department of Physics, Brown University, Providence, RI, 02912, USA}
\begin{abstract}
Loop corrections to observables in slow-roll inflation are found to
diverge no worse than powers of the log of the scale factor, extending
Weinberg's theorem to quasi-single field inflation models. Demanding
perturbation theory be valid during primordial inflation leads to
constraints on the effective lagrangian. This leads to some interesting
constraints and coincidences on the landscape of inflationary vacua.
\end{abstract}
\maketitle

\section{\textmd{\textup{Introduction}}}

In recent years there has been much discussion in the literature about
quantum effects of long wavelength modes in de Sitter, or slow-roll
inflationary backgrounds \cite{Weinberg:2005vy,Weinberg:2006ac,Polyakov:2007mm,Bartolo:2007ti,Enqvist:2008kt,Senatore:2009cf,Polyakov:2009nq,Burgess:2009bs,Higuchi:2010xt,Hollands:2010pr,Rajaraman:2010zx,Seery:2010kh,Krotov:2010ma,Marolf:2010nz,Marolf:2010zp,Hollands:2011we,Marolf:2011sh}.
Depending on the authors, these contributions are negligible, infinite,
or somewhere in between. A clear understanding of these issues is
therefore important in light of the experimentally verified predictions
of the semiclassical inflation theory. Essential to these predictions
is the assumption that the dominant contributions to density perturbations
are infrared finite, mode-by-mode.

In previous work \cite{Larjo:2011uh}, we emphasized the importance
of physical constraints on the choice of initial state and explained
how this leads to a theoretical uncertainty in the predictions for
the observations of a local observer. For example, in a global de
Sitter spacetime, perturbation theory in massive scalar field theory
around the Bunch-Davies vacuum appears convergent. Nevertheless it
is difficult to explicitly introduce an infrared cutoff, and then
remove it maintaining the symmetries. Depending on one's choice of
spacelike slices, such a procedure may be necessary. Moreover once
massless fields are included (even the graviton) the procedure of
adopting an infrared cutoff appears to fail, and it seems likely the
global spacetime is unstable.

On the other hand, for realistic applications to cosmology we are
more interested in a local patch of quasi-de Sitter spacetime that
expands to our observable universe. In this scenario a comoving infrared
cutoff is the simplest accurate model, and most of the questions of
principle for global de Sitter become irrelevant \cite{Larjo:2011uh,Xue:2011hm}.
In this context, any sensitivity of observables to the infrared cutoff
reflects a genuine theoretical uncertainty in predictions, originating
from the lack of a precisely controlled initial state. Such quantum
corrections were explored in \cite{Larjo:2011uh}.

In the present work our goal is to extend these results to slow-roll
inflation, allowing for the nontrivial time dependence of the Hubble
parameter. For models built using scalars with minimal kinetic terms,
it is straightforward to combine the physical setup of \cite{Larjo:2011uh}
with the results of Weinberg \cite{Weinberg:2005vy,Weinberg:2006ac}
for this class of models to see that observables at most diverge as
a power of a logarithm of the scale factor. While this presents serious
problems for the global stability of de Sitter spacetime, the infrared
quantum corrections are tiny for primordial slow-roll inflation with
realistic parameters.

Xue, Gao and Brandenberger \cite{Xue:2012wi} have proposed a related
scalar model with non-minimal kinetic terms that evade this conclusion.
They find large infrared quantum corrections produce strict bounds
on the scalar couplings arising from convergent perturbation theory.
In the present work we extend the results of Weinberg to this class
of nonminimal kinetic term models, and confirm that at most powers
of the logarithm of the scale factor appear in observables. We then
re-examine bounds on the couplings by requiring a good perturbative
expansion, and find that running of the scalar mass parameters tends
to produce an even larger effect than that of the infrared modes,
with somewhat less strict bounds emerging than found in \cite{Xue:2012wi}.
We also show that the infrared corrections in slow roll are bounded
above by the corresponding corrections in pure de Sitter spacetime,
as one would intuitively expect.

It is interesting to note that these bounds arising from quantum consistency
are not far off the kinds of bounds that emerge from tree-level slow-roll
considerations, combined with matching the scalar potential to the
magnitude of observed density fluctuations \cite{Bardeen:1983qw,Falk:1992zg}.
We argue this coincidence may be explained using statistics on a landscape
of vacuum states. Thus the saturation of the perturbative bound on
the landscape (at least within this class of models) may be regarded
a postdiction of the observed density fluctuations. We conclude with
a brief discussion of how the late-time instability of a de Sitter
region is compatible with the embedding of a de Sitter region in a
unitary model for quantum gravity \cite{Lowe:2007ek,Lowe:2010np}
and how the instability timescale that emerges solves the Boltzmann
brain paradox of cosmology \cite{Linde:2006nw}.

\section{\textmd{\textup{In-in formalism with IR cutoff}}}

We consider slow-roll inflation, with an infrared cutoff imposed as
in \cite{Larjo:2011uh}. To obtain a tractable model of slow-roll
inflation we consider a quasi-single field inflaton model with two
scalars: a slowly rolling inflaton $(\varphi)$ and a spectator field
$(\sigma)$, as already considered in \cite{Chen:2009zp} and \cite{Xue:2012wi}.
Using polar coordinates in field space, the inflaton and spectator
correspond to the tangential and radial directions respectively, and
the curvature of the inflaton trajectory leads to a minimal coupling
between the fields. The action governing the system is%
\footnote{This action differs form the one considered in \cite{Chen:2009zp}
in that our inflaton has been scaled by $R$ for later convenience,
$\varphi_{{\rm us}}=R\,\theta_{{\rm them}}$. Note that the kinetic
term for $\theta$ would be $-\frac{1}{2}(R+\sigma)^{2}g^{\mu\nu}\partial_{\mu}\theta\partial_{\nu}\theta$,
as expected when $\theta$ is the tangential coordinate and $(R+\sigma)$
is the distance from the origin. With our conventions, both scalar
fields have units of energy, as does $R$.%
}
\begin{equation}
S=\int d^{4}x\,\sqrt{-g}\left[-\frac{1}{2}\left(1+\frac{\sigma}{R}\right)^{2}g^{\mu\nu}\partial_{\mu}\varphi\partial_{\nu}\varphi-\frac{1}{2}g^{\mu\nu}\partial_{\mu}\sigma\partial_{\nu}\sigma-V(\varphi,\sigma)\right],\label{action1}
\end{equation}
where $R$ is a constant and the potential $V$ will be constrained
in such a way that slow-roll conditions for the inflaton are satisfied.
We will work in the spatially flat gauge, in which the metric is given
by 
\begin{equation}
ds^{2}=-dt^{2}+a(t)^{2}d\vec{x}^{2},
\end{equation}
and the scalar metric perturbation has been incorporated in the perturbation
of the inflaton field $\varphi$. It will also be convenient to use
the collective notation 
\begin{equation}
\vec{\Phi}=\left(\begin{array}{c}
\varphi\\
\sigma
\end{array}\right).
\end{equation}

\paragraph{\textup{Background solution:}}

Perturbing the fields via $\vec{\Phi}=\vec{\Phi}_{0}+\delta\vec{\Phi}$,
and then minimizing the action, we find the field equations governing
the background solution. Taking the background solution to be spatially
homogeneous, the background equations are 
\begin{align}
 & \ddot{\varphi}_{0}+3H\dot{\varphi}_{0}+V'_{\phi}=0,\label{background-phi}\\
 & \ddot{\sigma}_{0}+3H\dot{\sigma}_{0}+V'_{\sigma}-\frac{\dot{\varphi}_{0}^{2}}{R}=0,\label{background-sigma}
\end{align}
where $H\equiv\frac{\dot{a}}{a}$ is the Hubble parameter and $V'_{\varphi}\equiv\partial_{\varphi}V$
etc. Equation (\ref{background-phi}) places constraints on the potential
for the field $\varphi_{0}$ to undergo slow-roll. For the spectator
field we pick a constant solution, and without loss of generality
we can choose $\sigma_{0}=0$. Equation (\ref{background-sigma})
then relates the steepness of the potential in the radial direction
to the speed of rolling inflaton by
\begin{equation}
V'_{\sigma}=\frac{\dot{\varphi}_{0}^{2}}{R}\equiv R\lambda(t)^{2}.\label{lambda}
\end{equation}
Note the slow roll parameter is non-vanishing for non-zero $\lambda$
\begin{eqnarray}
\epsilon & \equiv & \frac{m_{pl}^{2}}{16\pi}\left(\frac{V'}{V}\right)^{2}\approx\frac{4\pi\dot{\varphi_{0}}^{2}}{H^{2}m_{pl}^{2}}=\frac{4\pi R^{2}\lambda^{2}}{H^{2}m_{pl}^{2}}\,.\label{eq:slowrollparam}
\end{eqnarray}

\subsection{\textmd{The free action}}

We wish to use in-in formalism to compute two-point correlators of
the form
\begin{equation}
G_{ij}(t)=\langle\left(Te^{-i\int_{-\infty_{-}}^{t}dt'H_{{\rm int}}}\right)^{\dagger}\delta\Phi_{i}(t)\,\delta\Phi_{j}(t)\left(Te^{-i\int_{-\infty_{+}}^{t}dt''H_{{\rm int}}}\right)\rangle,\label{two-point}
\end{equation}
where the mode functions $\delta\Phi$ are determined by the free
part of the Hamiltonian, and the interaction part is taken into account
perturbatively as in (\ref{two-point}). Thus we need the free Hamiltonian,
which is defined as the part quadratic in perturbations \cite{Adshead:2009cb,Weinberg:2005vy}.
Expanding the action around the background solution $\Phi_{0}$ up
to second order yields
\begin{align}
S_{{\rm free}} & =S_{0}+\frac{1}{2}\int d^{4}x\sqrt{-g}\left[(\dot{\delta\varphi})^{2}-a^{-2}(\nabla\delta\varphi)^{2}-V_{\varphi\varphi}''(\delta\varphi)^{2}+(\dot{\delta\sigma})^{2}-a^{-2}(\nabla\delta\sigma)^{2}\right.\nonumber \\
 & \quad\quad\left.-\left(V_{\sigma\sigma}''-\lambda^{2}\right)(\delta\sigma)^{2}+4\lambda\delta\sigma(\dot{\delta\varphi})-2V_{\sigma\varphi}''\delta\sigma\delta\varphi\right]\label{action-free}\\
 & =S_{0}+\frac{1}{2}\int d^{3}\vec{k}\, dt\,\sqrt{-g}\left[\dot{\varphi}_{k}^{2}-\left(\frac{k^{2}}{a^{2}}+V_{\varphi\varphi}''\right)\varphi_{k}^{2}+\dot{\sigma}_{k}^{2}-\left(\frac{k^{2}}{a^{2}}+V_{\sigma\sigma}''-\lambda^{2}\right)\sigma_{k}^{2}\right.\nonumber \\
 & \quad\quad\left.+4\lambda\sigma_{k}\dot{\varphi}_{k}-2V_{\sigma\varphi}''\sigma_{k}\varphi_{k}\right],\label{action-momentum}
\end{align}
where $S_{0}\equiv\int d^{4}x\sqrt{-g}[\frac{1}{2}\dot{\varphi}_{0}^{2}-V(\varphi_{0},\sigma_{0})]$
contains the zeroth-order terms, and we have switched to momentum
space via 
\begin{equation}
\delta\vec{\Phi}(t,\vec{x})=\int\frac{d^{3}\vec{k}}{(2\pi)^{\frac{3}{2}}}e^{i\vec{k}\cdot\vec{x}}\vec{\Phi}_{k}(t)\,.
\end{equation}
Note that in Fourier space we drop the $\delta$ in front of the perturbation.

\paragraph{\textup{The field equations:}}

From (\ref{action-momentum}) one can derive the field equations
\begin{align}
\ddot{\varphi}_{k}+3H\dot{\varphi}_{k}+\left(\frac{k^{2}}{a^{2}}+V_{\varphi\varphi}''\right)\varphi_{k} & =-2\partial_{t}\left(\lambda\sigma_{k}\right)-6H\lambda\sigma_{k}-V_{\sigma\varphi}''\sigma_{k}\,,\\
\ddot{\sigma}_{k}+3H\dot{\sigma}_{k}+\left(\frac{k^{2}}{a^{2}}+V_{\sigma\sigma}''-\lambda^{2}\right)\sigma_{k} & =2\lambda\dot{\varphi}_{k}\,.
\end{align}
We take the inflaton to be massless ($V''_{\varphi\varphi}=0$), and
denote by $m^{2}\equiv V''_{\sigma\sigma}-\lambda^{2}$ the effective
`mass' of the spectator field $\sigma$. We also take the potential
to be of the form $V=V(\varphi)+V(\sigma)$ to leading order, implying
$V''_{\varphi\sigma}=0$. Both of these constraints are consistent
with the analysis of \cite{Chen:2009zp,Xue:2012wi}.

In order to have a solvable system, from now on we will also take
the inflaton to roll at a constant speed, so $\dot{\lambda}=0$. Then
(\ref{background-sigma}) tells us that the potential has to be chosen
such that the slope $V'_{\sigma}$ is constant along the trajectory.
This departs from the analysis of \cite{Chen:2009zp,Xue:2012wi},
who make no such assumption. We choose to set this constraint, because
a central tenet of this article is that in order to compute correlators
of type (\ref{two-point}) one has to treat $\lambda$ analytically
in an exact manner, as opposed to perturbatively. In \cite{Chen:2009zp,Xue:2012wi}
the cross-term $\lambda\sigma\dot{\varphi}$ is bundled into the interaction
Hamiltonian, whereas we treat it as a part of $H_{{\rm free}}$, and
restrict to constant $\lambda$ in order to be able to explicitly
solve the field equations.

Finally, at the level of the field equations we will work in an `instantaneously
de Sitter' approximation, in which the scale factor is given by $a(t)=\exp(Ht)$,
with a constant $H$. This approximation is valid over time scales
\[
H^{2}\epsilon\Delta t\ll H\,\Longrightarrow\Delta t\ll\frac{Hm_{pl}^{2}}{4\pi R^{2}\lambda^{2}}
\]
which is quite sufficient for our purposes. In particular, it can
contain the regime where effects nonperturbative in $\lambda\Delta t$
become important. Such effects are dropped in \cite{Chen:2009zp,Xue:2012wi}
where a perturbative expansion is $\lambda$ is considered.

At this point it is also convenient to switch to conformal time, defined
by

\begin{equation}
dt=a(t)d\tau,\quad\Rightarrow\quad\tau=\int\frac{dt}{a(t)}=-\frac{1}{Ha}.
\end{equation}
Incorporating the constraints and approximations the field equations
in conformal time become 
\begin{align}
 & \varphi''_{k}-\frac{2}{\tau}\varphi'_{k}+k^{2}\varphi_{k}=\frac{2\lambda}{H\tau}\left(\sigma'_{k}-\frac{3}{\tau}\sigma_{k}\right),\label{fieldeq-phi}\\
 & \sigma''_{k}-\frac{2}{\tau}\sigma'_{k}+\left(k^{2}+\frac{m^{2}}{H^{2}\tau^{2}}\right)\sigma_{k}=-\frac{2\lambda}{H\tau}\varphi'_{k},\label{fieldeq-sigma}
\end{align}
 where $'\equiv\partial_{\tau}$.

\subsection{\textmd{The free field solution}}

We will now solve the field equations (\ref{fieldeq-phi},\ref{fieldeq-sigma})
perturbatively in $k$ using the Green's function method. Using the
expansion 
\begin{equation}
\vec{\Phi}=\sum_{i=0}^{\infty}k^{2i}\vec{\Phi}_{i}\,,
\end{equation}
we can write the field equations in matrix notation as 
\begin{equation}
L\vec{\Phi}_{i}=-k^{2}\vec{\Phi}_{i-1},\quad\quad{\rm with}\quad L\equiv\left(\begin{array}{cc}
\partial_{\tau}^{2}-\frac{2}{\tau}\partial_{\tau} & -\frac{2\lambda}{H\tau}\left(\partial_{\tau}-\frac{3}{\tau}\right)\\
\frac{2\lambda}{H\tau}\partial_{\tau} & \partial_{\tau}^{2}-\frac{2}{\tau}\partial_{\tau}+\frac{m^{2}}{H^{2}\tau^{2}}
\end{array}\right).\label{fieldeq-i}
\end{equation}
 Note that one should not confuse the mode function $\vec{\Phi}_{i=0}$
with the values of the background fields $\vec{\Phi}_{0}$ found earlier.
From now on the background values will only appear inside $\lambda=\dot{\varphi}_{0}/R$,
so no confusion should arise.

We can easily solve $\vec{\Phi}_{0}$ from $L\vec{\Phi}_{0}=0$, which
has power law solutions. One verifies that the general solution is
\begin{align}
\vec{\Phi}_{0} & =\sum_{i=1}^{4}\vec{A}_{i}\left(\frac{\tau}{\tau_{0}}\right)^{\alpha_{i}},\quad{\rm with}\quad\vec{\alpha}=(0,3,\alpha_{-},\alpha_{+}),\quad\vec{A}_{i}=\left(\begin{array}{c}
a_{i}^{\varphi}\\
a_{i}^{\sigma}
\end{array}\right),\label{phi-power-sol}\\
 & \alpha_{\pm}=\frac{3}{2}\left(1\pm\sqrt{1-\left(\frac{2m}{3H}\right)^{2}-\left(\frac{4\lambda}{3H}\right)^{2}}\right),\nonumber \\
 & \vec{a}^{\sigma}=\left(0\,\,,\,\,\,-\frac{6\lambda H}{m^{2}}a_{2}^{\varphi}\,\,,\,\,\,\frac{3+\alpha_{-}-\alpha_{+}}{4\lambda/H}\, a_{3}^{\varphi}\,\,,\,\,\,\frac{3-\alpha_{-}+\alpha_{+}}{4\lambda/H}\, a_{4}^{\varphi}\,\,\right),\nonumber 
\end{align}
where $\tau_{0}$ is a fixed initial time. We also need the Green's
function, defined by 
\begin{equation}
LG(\tau,\tau')=\delta(\tau-\tau')\,\mathbbm{1},\quad{\rm with}\quad G=\left(\begin{array}{cc}
G_{\varphi\varphi} & G_{\varphi\sigma}\\
G_{\sigma\varphi} & G_{\sigma\sigma}
\end{array}\right).\label{green-def}
\end{equation}
 We relegate the computation of $G$ into appendix \ref{app-green},
here we only present the result 
\begin{equation}
G(\tau,\tau')=\sum_{i=1}^{4}\bar{C}_{i}\,\Theta(\tau'-\tau)\,\tau'\,\left(\frac{\tau}{\tau'}\right)^{\alpha_{i}},
\end{equation}
where the $\bar{C}_{i}$ are constant matrices explicitly given in
(\ref{Cbars}).

We can now use the Green's function iteratively to solve for higher
orders $\Phi_{i}$. We have 
\begin{equation}
\vec{\Phi}_{1}(\tau)=-k^{2}\int_{\tau_{0}}^{0}d\tau'\, G(\tau,\tau')\cdot\vec{\Phi}_{0}(\tau')=\left(k\tau_{0}\right)^{2}\sum_{i,j=1}^{4}\frac{\bar{C}_{i}\cdot\vec{A}_{j}}{2-\alpha_{i}+\alpha_{j}}\tau_{0}^{\alpha_{j}}\left(\frac{\tau}{\tau_{0}}\right)^{2+\alpha_{j}}.\label{eq:phicorrect}
\end{equation}

\paragraph{\textup{Late times:}}

The four independent solutions at late-time take the form
\begin{eqnarray}
\vec{\Phi}(\tau) & = & a_{0}^{\varphi}\left(\begin{array}{c}
1+(k\tau_{0})^{2}a_{0,1}^{\varphi}\left(\frac{\tau}{\tau_{0}}\right)^{2}+\cdots\\
(k\tau_{0})^{2}a_{0,1}^{\sigma}\left(\frac{\tau}{\tau_{0}}\right)^{2}+\cdots
\end{array}\right)+a_{1}^{\varphi}\left(\begin{array}{c}
\left(\frac{\tau}{\tau_{0}}\right)^{3}+(k\tau_{0})^{2}a_{0,1}^{\varphi}\left(\frac{\tau}{\tau_{0}}\right)^{5}+\cdots\\
-\frac{6\lambda H}{m^{2}}\left(\frac{\tau}{\tau_{0}}\right)^{3}+(k\tau_{0})^{2}a_{1,1}^{\sigma}\left(\frac{\tau}{\tau_{0}}\right)^{5}+\cdots
\end{array}\right)\nonumber \\
 & + & a_{2}^{\varphi}\left(\begin{array}{c}
\left(\frac{\tau}{\tau_{0}}\right)^{\alpha_{-}}+(k\tau_{0})^{2}a_{2,1}^{\varphi}\left(\frac{\tau}{\tau_{0}}\right)^{\alpha_{-}+2}+\cdots\\
\frac{3+\alpha_{-}-\alpha_{+}}{4\lambda/H}\left(\frac{\tau}{\tau_{0}}\right)^{\alpha_{-}}+(k\tau_{0})^{2}a_{2,1}^{\sigma}\left(\frac{\tau}{\tau_{0}}\right)^{\alpha_{-}+2}+\cdots
\end{array}\right)\nonumber \\
 & + & a_{3}^{\varphi}\left(\begin{array}{c}
\left(\frac{\tau}{\tau_{0}}\right)^{\alpha_{+}}+(k\tau_{0})^{2}a_{3,1}^{\varphi}\left(\frac{\tau}{\tau_{0}}\right)^{\alpha_{+}+2}+\cdots\\
\frac{3-\alpha_{-}+\alpha_{+}}{4\lambda/H}\left(\frac{\tau}{\tau_{0}}\right)^{\alpha_{+}}+(k\tau_{0})^{2}a_{3,1}^{\sigma}\left(\frac{\tau}{\tau_{0}}\right)^{\alpha_{+}+2}+\cdots
\end{array}\right)\label{eq:latetime}
\end{eqnarray}
where the new coefficients $a_{i,1}^{\varphi}$ and $a_{i,1}^{\sigma}$
are functions only of $m,\lambda$ and may be read-off from \eqref{eq:phicorrect}.
The terms $\cdots$ denote subleading terms as $\tau\to0$.

\subsection{Quantization}

We quantize these fields using the mode expansion
\[
\Phi(x,t)=\int d^{3}q\, e^{iq\cdot x}\Phi(q,t)\cdot\boldsymbol{\mathbf{\alpha}}(q)+e^{-iq\cdot x}\Phi^{*}(q,t)\cdot\boldsymbol{\alpha}^{*}(q)\,,
\]
where
\[
[\alpha_{i}(q),\alpha_{j}^{*}(q')]=\delta_{ij}\delta^{3}(q-q')\,,\qquad[\alpha_{i}(q),\alpha_{j}(q')]=0\,,
\]
and take the vacuum to be the Bunch-Davies vacuum at the start of
inflation, annihilated by these annihilation operators. In this early
time limit, the modes of interest (recall the comoving infrared cutoff)
are all inside the horizon, and oscillate with time. The quantization
proceeds in the standard way.

Later we will need to also estimate the commutator of the fields in
the late-time limit, where the fields asymptote to the form \eqref{eq:latetime}.
Now the modes of interest are far outside the horizon where they decay
as real powers of the scale factor \eqref{eq:latetime}. At leading
order $ $as $\tau\to0$ (late times), 
\[
\Phi(q,t)=\left(\begin{array}{cc}
C_{q}\tau^{0}+D_{q}\tau^{3} & E_{q}\tau^{\alpha_{-}}+F_{q}\tau^{\alpha_{+}}\\
-\frac{6\lambda H}{m^{2}}D_{q}\tau^{3} & \frac{3+\alpha_{-}-\alpha_{+}}{4\lambda/H}E_{q}\tau^{\alpha_{-}}+\frac{3-\alpha_{-}+\alpha_{+}}{4\lambda/H}F_{q}\tau^{\alpha_{+}}
\end{array}\right)
\]
where the complex coefficients $C_{q},D_{q},E_{q},F_{q}$ are fixed
by matching to the early time modes at horizon crossing. When we compute
the commutator $[\Phi_{i}(x,t),\Phi_{j}(x',t')]$ only cross terms
between the pairs $\tau^{0},\tau^{3}$ and $\tau^{\alpha_{-}},\tau^{\alpha_{+}}$
survive, so the commutator falls off as $\tau^{3}$ as $\tau\to0$.

\subsection{Late-time limit of observables}

We follow Weinberg \cite{Weinberg:2005vy,Weinberg:2006ac} when computing
the leading late-time terms. As emphasized in his work, there are
delicate cancellations between terms, which are most easily taken
care of using the commutator expression
\begin{equation}
\left\langle Q(t)\right\rangle =\sum_{N=0}^{\infty}i^{N}\int_{t_{0}}^{t}dt_{N}\int_{t_{0}}^{t_{N}}dt_{N-1}\cdots\int_{t_{0}}^{t_{2}}dt_{1}\left\langle \left[H_{I}(t_{1}),\left[H_{I}(t_{2}),\cdots\left[H_{I}(t_{N}),Q_{I}(t)\right]\cdots\right]\right]\right\rangle \label{eq:commutator}
\end{equation}
where the subscript $I$ denotes interaction picture operators. Note
that rather than taking $t_{0}\to-\infty$ as in \cite{Weinberg:2005vy,Weinberg:2006ac},
we keep it finite, as part of the procedure introduced in \cite{Larjo:2011uh}
for keeping track of the effect of the initial state on observables.
Weinberg investigated the leading late-time divergences for massless
and massive minimally coupled scalar fields, as well as Dirac particles
and vector particles. Here we will apply this approach to quasi-single
field inflation.

As argued above, the commutator of any pair of elementary fields falls
off as $\tau^{3}$ or $a(t)^{-3}$. The same will also be true if
one considers time or space derivatives of these fields. The interaction
hamiltonian $H_{I}$ will be built of products of such fields and
derivatives, and contain a volume derivative going as $a(t)^{3}$.
As is apparent from \eqref{eq:commutator} the $H_{I}$ will always
appear inside a commutator, so these factors of $a(t)$ will cancel.

In general observable we may also encounter additional powers of fields
that are not inside commutators. These introduce at worst constant
factors (if only $\varphi(x,t)$ appears in $Q(t)$) or factors that
fall at least as fast as $\tau^{\alpha_{-}}$, provided some factors
of $\sigma(x,t)$ appear, or provided sufficient inverse powers of
$a(t)$ appear in derivatives. 

We conclude then that observables built solely out of products of
$\varphi(x,t)$ do have late time divergences, arising from the time
integrals in \eqref{eq:commutator}. These lead to divergences as
powers of $\log\tau$. These divergences are qualitatively the same
as the case of an interacting massless minimally coupled scalar considered
in \cite{Larjo:2011uh}. Of course, in order for inflation to end,
this scalar must acquire a mass, which provides a natural physical
cutoff to these time integrals. Estimates made in \cite{Larjo:2011uh}
(in the models considered there) show these loop effects are negligible
for primordial inflation compared to the tree-level contributions,
and only become significant in genuine asymptotic future de Sitter
phases -- where they are capable of driving an instability. Nevertheless
there exists a large class of infrared finite slow-roll observables
for quasi-single field inflation which contain either derivatives
of $\varphi(x,t)$ or factors of $\sigma(x,t)$.

\section{Loop corrections and renormalization: infrared and uv contributions}

The authors of \cite{Xue:2012wi} considered a closely related set
of questions, and argued much larger infrared terms appear due to
slow-roll effects. They find divergences that go like powers of $\left(\frac{H_{initial}}{H}\right)^{2}/\epsilon$
(with $\epsilon$ the slow-roll parameter \eqref{eq:slowrollparam},
$H$ the late-time Hubble parameter, and $H_{initial}$ the Hubble
parameter at the start of inflation). Here we will show these conclusions
change when UV divergences are treated with a physical renormalization
prescription. The apparent divergences as $\epsilon\to0$ disappear,
but are replaced by late time divergences involving powers of $\log a(t)$.
Such divergences match those expected from the de Sitter limit. Demanding
that these late-time loop corrections not destroy perturbation theory
leads to constraints on the interaction potential, which take a similar
form to those argued for in \cite{Xue:2012wi}, but differ in the
details.

In section III.1 of \cite{Xue:2012wi} they consider loop corrections
to the scalar curvature perturbation two-point function (we refer
to \cite{Xue:2012wi} for details), due to a massless (or sufficiently
light) entropy perturbation. This can be represented by the $\sigma$
field of the model previously discussed, with a higher order $ $$g\sigma^{4}$
self-coupling included. The result of their analysis is that the leading
IR divergences comes from a subdiagram involving the $\sigma$ self-interaction
\begin{figure}
\includegraphics[scale=1.5]{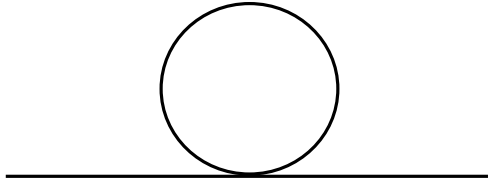}\caption{\label{fig:Infrared-divergent-diagram}Infrared divergent diagram}
\end{figure}
. The corrections appears in eqn (56) \cite{Xue:2012wi}. Carrying
over their result for the subdiagram, and also including a physical
UV cutoff we obtain
\begin{eqnarray}
g\int_{\Lambda_{IR}}^{\Lambda_{IR,phys}a(t)}\frac{dq}{q}H_{initial}^{2}\left(\frac{q}{k_{initial}}\right)^{-2\epsilon}+g\int_{\Lambda_{IR,phys}}^{\Lambda_{UV,phys}}\frac{d^{3}q}{q^{2}} & = & \frac{gH_{initial}^{2}}{2\epsilon}\left(1-e^{-2\epsilon H_{initial}t}\left(\frac{\Lambda_{IR,phys}}{H_{initial}}\right)^{-2\epsilon}\right)\nonumber \\
 & + & g\left(\Lambda_{UV,phys}^{2}-\Lambda_{IR,phys}^{2}\right)\,,\label{eq:massterm}
\end{eqnarray}
where we have estimated the UV and IR divergent terms by breaking
the range of integration up at a physical intermediate scale, and
used the appropriate asymptotic forms of the propagator. Note a comoving
IR cutoff and a proper UV cutoff is used as in \cite{Larjo:2011uh}.
The computation of \cite{Xue:2012wi} also uses a comoving IR cutoff,
but are less explicit about their choice of UV cutoff. A mass counterterm
must be chosen to cancel the UV divergence, and impose a renormalization
condition. This is described in more detail in appendix \ref{sec:Renormalization-in-cosmological}.
The result is that the would-be $H_{initial}^{2}/\epsilon$ divergence
disappears when a physical renormalization prescription is imposed
for the mass of the $\sigma$ field.

It is also worth pointing out that even the long wavelength contribution
to \eqref{eq:massterm} is bounded from above by the pure de Sitter
result, where $\epsilon=0$. This follows by choosing the scale $k_{initial}=\Lambda_{IR}=\mathcal{O}(H_{initial})$,
and noting for slow-roll with $\epsilon>0$, the integrand is always
positive and less than the pure de Sitter answer ($\epsilon=0)$ throughout
the range of integration. This point is at odds with the answer obtained
in \cite{Xue:2012wi}, which may be traced to them dropping all but
the first term on the right-hand side of \eqref{eq:massterm}.

An important consistency condition is demanding that the perturbative
expansion in $g$ converge. In analyzing this question, we allow for
the mass renormalization of the inflaton to incorporate generic effects
of new physics near the GUT scale, and still require the perturbative
expansion be valid. The implications for the one-loop renormalization
are described in appendix \ref{sec:Renormalization-in-cosmological}.
Let us examine the physical constraints that emerge from this. If
we insert this subdiagram into the full expression for the one loop
correction to the two-point function \cite{Larjo:2011uh} (simply
working in the $g\sigma^{4}$ sector of the theory), and ask when
perturbation theory is valid, we obtain the condition at the end of
inflation that
\[
gH^{2}\left(N^{3}+N^{2}\frac{\Lambda_{GUT}^{2}}{H^{2}}\right)\ll H^{2}(N-\log\left(-\Lambda_{IR}\tau_{0}\right))\,.
\]
Now the $N^{3}$ term on the left yields a constraint $g\ll1/N^{2}\sim10^{-4}$
for massless perturbation theory to be valid. The other term requires
\[
gN\frac{\Lambda_{GUT}^{2}}{H^{2}}\ll1\,.
\]
Let us put in some typical values assuming we wish to use the field
theory for the inflaton from a UV cutoff near the GUT scale (beyond
which we expect new physics to set in), down to scales below the Hubble
scale. We set $H_{initial}=10^{14}GeV$ and $\Lambda_{GUT}=10^{16}GeV$.
This yields
\[
g\ll10^{-6}\,.
\]

So we see while these effects are compatible with the bounds of \cite{Xue:2012wi}
the bounds found there are not the whole story, and stronger constraints
emerge from the consideration of typical UV effects due to renormalization.
Another difference with \cite{Xue:2012wi} is the powers of $N$.
In both our work, and \cite{Xue:2012wi} a comoving infrared cutoff
is used. The justification for this is elaborated in \cite{Larjo:2011uh}.
However the computation of \cite{Xue:2012wi} appears to use estimates
for amplitudes obtained using the formalism of \cite{Burgess:2009bs},
who instead use a physical/proper distance infrared cutoff.

\begin{figure}
\includegraphics[scale=0.75]{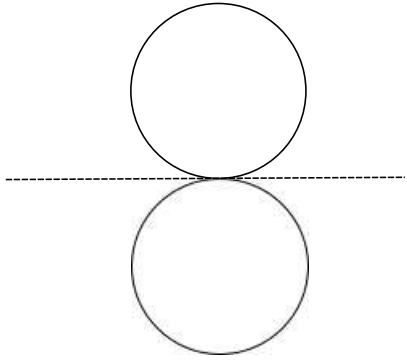}

\caption{Sub-diagram with two scalar curvature external lines, and the scalar
field lines fully contracted.}
\end{figure}

Finally we can estimate the two loop contribution coming from the
diagram shown in figure 2, arising from the coupling of the scalar
curvature to the scalar field as described in \cite{Xue:2012wi}.
Following the same type of computation as above, we find a constraint
of the form
\[
gH^{2}\left(N^{4}+N^{3}\left(\frac{\Lambda_{GUT}^{2}}{H^{2}}\right)+N^{2}\left(\frac{\Lambda_{GUT}^{2}}{H^{2}}\right)^{2}\right)\ll H^{2}(N-\log\left(-\Lambda_{IR}\tau_{0}\right))\,.
\]
This then yields the dominant condition
\begin{equation}
gN\frac{\Lambda_{GUT}^{4}}{H^{4}}\ll1\,,\label{eq:finalbound}
\end{equation}
so that $g\ll10^{-10}$ which numerically is comparable with the bound
of \cite{Xue:2012wi}, though the dominant effect is the short distance
renormalization of the scalar mass, rather than large distance slow-roll
terms.

\section{Comments on the Landscape }

If we had considered single-field inflation with a potential $g\sigma^{4}$,
a tree-level bound on $g$ emerges by matching the observed $\delta\rho/\rho\sim10^{-5}$
with the value predicted by slow-roll \cite{Bardeen:1983qw,Falk:1992zg}.
This yields $\delta\rho/\rho\gtrsim Ng^{1/2}$ so that $g<10^{-13}$.
It is interesting to point out that the loop-level bound \eqref{eq:finalbound}
is comparable with this tree-level bound. This coincidence suggests
an anthropic relation. Namely anthropic/landscape considerations would
tend to statistically favor the largest value of $g$ compatible with
the basic physics of the model. To make $g$ exceed our quantum bound
requires new physics to appear before the GUT scale, taking us out
of this class of model. Taking $g$ to saturate the bound, and assuming
for the sake of argument that one is restricted to working with this
family of scalar models, one is then led to a postdiction for $\delta\rho/\rho$
matching observation.%
\footnote{It is also worth pointing out a similar mechanism may be at work in
the Standard Model. With the measured mass of the Higgs boson at $125$
GeV, the Higgs potential develops an instability due to renormalization
group flow at an high scale, below the Planck scale \cite{Degrassi:2012ry}.
This hints that landscape statistics push the parameters of the Higgs
potential to a point of marginal stability before new physics takes
over.%
}

Finally, it is also worth commenting further on the gravitational
version of the late-time instability of de Sitter spacetime found
in \cite{Larjo:2011uh} due to the choice of graviton initial state.
It was argued there that future eternal de Sitter is actually unstable
on a timescale of $10^{122}$ e-folds. It has already been noted that
some kind of late time instability for de Sitter is needed for compatibility
with the class of unitary models for quantum de Sitter regions, considered
in \cite{Alberghi:1999kd,Lowe:2007ek,Lowe:2010np}. There is was pointed
out that this can solve the proliferation of Boltzmann brain observers
(see \cite{Linde:2006nw} for background material). 

In the present context, we find the timescale associated with the
production of a Boltzmann brain along the path of some timelike geodesic
in an expanding universe to be of order its inverse Boltzmann factor
$e^{E/kT}\approx e^{10^{65}}$ for an observer of order 1 mole of
protons, with $T$ the temperature of the present cosmological horizon.
This timescale is much larger than the above instability timescale.
The infrared instability of de Sitter spacetime thus has a chance
to restore our status as typical observers, solving one of the many
problems associated with doing statistics on a landscape of theory
vacua.
\begin{acknowledgments}
We thank R. Brandenberger for helpful comments. This research is supported
in part by DOE grant DE-FG02-91ER40688-Task A and an FQXi grant.
\end{acknowledgments}
\appendix

\section{\textmd{\textup{\label{sec:Renormalization-in-cosmological}Renormalization
in cosmological spacetimes}}}

It is helpful to review renormalization in the context of the models
considered here, and in \cite{Larjo:2011uh}, filling in some additional
details omitted in the earlier work. The scalar coupling $\lambda$
of \cite{Larjo:2011uh} will be replaced by $g$ here, to avoid confusion
with the discussion in Section 2.The loop diagram figure \ref{fig:Infrared-divergent-diagram}
gives rise to the integral of eqn. (9) of \cite{Larjo:2011uh}. The
potentially divergent terms arise from the IR and UV ends of the integral,
and take the form
\begin{eqnarray*}
L(\tau_{v}) & = & \frac{-ig}{(2\pi)^{2}H^{2}\tau_{v}^{4}}\left(\tau_{v}^{2\gamma}\int_{\Lambda_{IR}}^{-\tau_{v}^{-1}}dp\, p^{-1+2\gamma}+\tau_{v}^{2}\int_{-\tau_{v}^{-1}}^{\Lambda_{UV}a(\tau_{v})}dp\frac{p^{2}}{\sqrt{p^{2}+m^{2}a(\tau_{v})^{2}}}\right)\\
 & = & \frac{-ig}{(2\pi)^{2}H^{2}\tau_{v}^{4}}\left(\frac{1-\left(-\Lambda_{IR}\tau_{v}\right)^{2\gamma}}{2\gamma}+\frac{1}{2}\left(\frac{\Lambda_{UV}}{H}\right)^{2}-\frac{1}{2}\left(\frac{m}{H}\right)^{2}\log\frac{\Lambda_{UV}}{m}\right)\,,
\end{eqnarray*}
with $\gamma=m^{2}/(3H^{2})$, and $\tau_{v}$ the conformal time
of the vertex factor insertion. Here we have included the subleading
log UV divergent term omitted in \cite{Larjo:2011uh}, and assumed
$\Lambda_{UV}\gg m$. 

Now let us consider choosing the counter-term so that a physical mass
renormalization condition is imposed at some scale $\mu$. In keeping
with the small mass/early time expansion used in this paper, we expand
the IR term in powers of $m^{2}$ to get 
\[
L(\tau_{v})=\frac{ig}{2(2\pi)^{2}H^{2}\tau_{v}^{4}}\left(\log\left(-\Lambda_{IR}\tau_{v}\right)-\left(\frac{\Lambda_{UV}}{H}\right)^{2}+\left(\frac{m}{H}\right)^{2}\log\frac{\Lambda_{UV}}{m}\right)\,.
\]
In this way, we see that with a comoving infrared cutoff, we cannot
completely eliminate the IR divergence into a mass renormalization,
due to the additional $\log\tau_{v}$ dependence, which causes problems
at very late times. With a proper IR cutoff such a renormalization
is possible, as discussed in \cite{Burgess:2009bs}, however as discussed
in \cite{Larjo:2011uh} such proper IR cutoffs are unphysical for
spacetimes of cosmological interest (an example being bubble walls
moving faster than the speed of light).

Now a mass counter-term produces a shift
\[
\delta L(\tau_{v})=-\frac{i}{H^{4}\tau_{v}^{4}}\delta m^{2}\,,
\]
so comparing the UV divergent terms with what we usually have with
flat spacetime renormalization we choose to impose the renormalization
condition
\[
\delta m^{2}+\frac{g}{8\pi^{2}}\left(\Lambda_{UV}^{2}-m^{2}\log\frac{\Lambda_{UV}}{m}-H^{2}\log\Lambda_{IR}\right)=m_{phys}^{2}\,.
\]
Now we wish to impose the renormalization group equation $\Lambda_{UV\,}dm_{phys}^{2}/d\Lambda_{UV}=0$
and view $m^{2}+\delta m^{2}$ as the bare mass squared $m_{0}^{2}$,
which implies
\[
\Lambda_{UV}\frac{dm_{0}^{2}}{d\Lambda_{UV}}+\frac{g}{8\pi^{2}}\left(2\Lambda_{UV}^{2}-m_{0}^{2}\right)=0\,.
\]
Integrating this equation we find
\[
m_{0}^{2}(\Lambda_{UV})=\frac{g}{8\pi}\left(m_{0}^{2}\log\Lambda_{UV}-\Lambda_{UV}^{2}\right)+c\,,
\]
where $c$ is a constant independent of $\Lambda_{UV}$ to be fixed
by the renormalization condition. Substituting we find the physical
mass
\[
m_{phys}^{2}=c+\frac{g}{8\pi^{2}}\left(-H^{2}\log\Lambda_{IR}+m_{0}^{2}\log m_{phys}\right)\,,
\]
therefore we fix
\[
c=m_{phys}^{2}-\frac{g}{8\pi^{2}}\left(-H^{2}\log\Lambda_{IR}+m_{phys}^{2}\log m_{phys}\right)\,,
\]
at leading order in $g$. Thus the loop diagram with mass counterterm
gives
\[
L(\tau_{v})=\frac{ig}{2(2\pi)^{2}H^{2}\tau_{v}^{4}}\log\left(-\tau_{v}\right)\,,
\]
in this light scalar/early time approximation, and all dependence
on the UV and IR cutoffs disappears, with the exception of the $\log(-\tau_{v})$
term which comes from the choice of comoving IR cutoff. Note that
additional dependence on the IR cutoff appears when the time integrals
of the in-in formulation are performed, as elaborated in \cite{Larjo:2011uh}.

Finally, it is useful to reconsider the above computation, assuming
instead that new physics sets in at some high physical scale $\Lambda_{GUT}>H$.
For example, a new field $\varphi$ of mass $\Lambda_{GUT}$ coupling
via $g\sigma^{2}\varphi^{2}$. In this case we do indeed find a correction
of the form
\[
L(\tau_{v})=\frac{i\lambda}{2(2\pi)^{2}H^{2}\tau_{v}^{4}}\left(\log\left(-\tau_{v}\right)-\left(\frac{\Lambda_{GUT}}{H}\right)^{2}\right)\,,
\]
showing new physics does indeed lead to a quadratic shift in the mass.
This form will be useful for estimating the range of the perturbative
validity of slow roll theory for energy scales approaching the GUT
scale.

\section{\textmd{\textup{Green's Function}}}

\label{app-green}In this appendix we solve equation (\ref{green-def})
to derive the Green's function of the system. Since we know the zeroth
order solution (\ref{phi-power-sol}), a good ansatz is 
\begin{equation}
G(\tau,\tau')=\sum_{k=1}^{4}C_{k}\Theta(\tau'-\tau)\tau^{\alpha_{k}},\quad{\rm with}\quad C_{k}=\left(\begin{array}{cc}
c_{11,k} & c_{12,k}\\
c_{21,k} & c_{22,k}
\end{array}\right),
\end{equation}
where the coefficients $c_{ij,k}$ satisfy the same relations as $a_{k}^{\varphi,\sigma},$
i.e. 
\begin{equation}
\frac{a_{k}^{\sigma}}{a_{k}^{\varphi}}=\frac{c_{21,k}}{c_{11,k}}=\frac{c_{22,k}}{c_{12,k}}.\label{c-cond1}
\end{equation}
 In order to fix the rest of the coefficients $c_{ij,k}$ we integrate
(\ref{green-def}) over the range $\tau\in[\tau'-\epsilon,\tau'+\epsilon]$,
computing to order $\mathcal{O}(\epsilon^{0})$, 
\begin{equation}
\mathbbm{1}=\int_{\tau'-\epsilon}^{\tau'+\epsilon}d\tau L\Theta(\tau-\tau')C_{k}\tau^{\alpha_{k}}=-\int_{\tau'-\epsilon}^{\tau'+\epsilon}d\tau\left(C_{k}\tau^{\alpha_{k}}\right)\frac{\delta(\tau-\tau')}{\tau-\tau'}+2\tau'^{\alpha_{k}}\left(\begin{array}{cc}
\alpha_{k}-1 & -\frac{\lambda}{H}\\
\frac{\lambda}{H} & \alpha_{k}-1
\end{array}\right)\cdot C_{k},
\end{equation}
where summing over repeated indices is implied. The first term arises
from $\partial_{\tau}^{2}\Theta(\tau'-\tau)$ and is potentially divergent;
we need to demand 
\begin{equation}
C_{k}\tau'^{\alpha_{k}}=0\label{c-cond2}
\end{equation}
for it to vanish. The second term implies 
\begin{equation}
2\tau'^{\alpha_{k}}\left(\begin{array}{cc}
\alpha_{k}-1 & \frac{\lambda}{H}\\
-\frac{\lambda}{H} & \alpha_{k}-1
\end{array}\right)\cdot C_{k}=\mathbbm{1}.\label{c-cond3}
\end{equation}
 Expressing $c_{2j,k}$ in terms of $c_{1j,k}$ using (\ref{c-cond1})
leaves us with eight unfixed coefficients ($c_{1j,k}$). The remaining
constraints (\ref{c-cond2}) and (\ref{c-cond3}) do not mix $c_{11,k}$
and $c_{12,k}$, and hence we are left with two groups of four unfixed
coefficients, with four constraints for each group. Hence solving
for $c_{ij,k}$ amounts to inverting $4\times4$ matrices, and we
find the coefficients to be given by
\begin{eqnarray}
C_{k} & \equiv & \bar{C}_{k}\tau'^{1-\alpha_{k}},\nonumber \\
\bar{C}_{1} & = & \left(\begin{array}{cc}
-\frac{m^{2}}{6(m^{2}+4\lambda^{2})} & \frac{\lambda}{m^{2}+4\lambda^{2}}\\
0 & 0
\end{array}\right),\qquad\bar{C}_{2}=\left(\begin{array}{cc}
\frac{m^{2}}{6(m^{2}+4\lambda^{2})} & 0\\
-\frac{\lambda}{m^{2}+4\lambda^{2}} & 0
\end{array}\right),\nonumber \\
\bar{C}_{3} & = & -\left(\begin{array}{cc}
\frac{2\lambda^{2}}{(m^{2}+4\lambda^{2})\sqrt{9-4m^{2}-16\lambda^{2}}} & \frac{2\lambda}{4m^{2}+16\lambda^{2}-9+3\sqrt{9-4m^{2}+16\lambda^{2}}}\\
\frac{\lambda\left(3-\sqrt{9-4m^{2}-16\lambda^{2}}\right)}{2(m^{2}+4\lambda^{2})\sqrt{9-4m^{2}-16\lambda^{2}}} & \frac{1}{2\sqrt{9-4m^{2}-16\lambda^{2}}}
\end{array}\right),\nonumber \\
\bar{C}_{4} & = & \left(\begin{array}{cc}
\frac{2\lambda^{2}}{(m^{2}+4\lambda^{2})\sqrt{9-4m^{2}-16\lambda^{2}}} & \frac{2\lambda}{9-4m^{2}-16\lambda^{2}+3\sqrt{9-4m^{2}+16\lambda^{2}}}\\
\frac{\lambda\left(3+\sqrt{9-4m^{2}-16\lambda^{2}}\right)}{2(m^{2}+4\lambda^{2})\sqrt{9-4m^{2}-16\lambda^{2}}} & \frac{1}{2\sqrt{9-4m^{2}-16\lambda^{2}}}
\end{array}\right),\label{Cbars}
\end{eqnarray}
where we set $H=1$; it can be restored by scaling $m\to\frac{m}{H}$
and $\lambda\to\frac{\lambda}{H}$.

\bibliographystyle{apsrev}
\bibliography{desittir}

\end{document}